\documentstyle[12pt]{article}
\font\twelvemsb=msbm10 scaled 1200
\def\Bbb#1{\hbox {\twelvemsb#1}}

\newcommand\ov[1]{\overline{#1}}
\newcommand{\M}{{\cal M}}
\newcommand{\Mg}{(\M,{}^4g)}
\newcommand{\N}{{\cal N}}
\newcommand{\V}{{\cal V}}

\newcommand{\HH}{{\cal H}}

\newcommand{\na}{\widehat{\nabla}}

\newtheorem{theorem}{Theorem}
\newtheorem{lemma}{Lemma}
\newtheorem{definition}{Definition}

\title{On uniqueness of static Einstein-Maxwell-dilaton black holes}
\author{Marc Mars\footnotemark[4] \, and Walter Simon\footnotemark[2]
\thanks{supported by Fonds zur F\"orderung der wissenschaftlichen
Forschung, project P14621-Mat.}  \\
\footnotemark[4] Albert Einstein Institut, \\
Am M\"uhlenberg 1, D-14476 Golm, Germany.\\
\footnotemark[2] Institut f\"ur Theoretische Physik der Univ. Wien \\
Boltzmanngasse 5, A-1090, Wien, Austria}
\begin{document}
\maketitle
\begin{abstract}
We prove uniqueness of static, asymptotically flat spacetimes with non-degenerate black
holes for three special cases of Einstein-Maxwell-dilaton theory: For the coupling ``$\alpha = 1$''
(which is the low energy limit of string theory) on the one  hand, and for vanishing magnetic or
vanishing electric field (but arbitrary  coupling) on the other hand. Our work generalizes in a natural,
but non-trivial way the uniqueness result obtained by Masood-ul-Alam who requires both $\alpha = 1$
and absence of magnetic fields, as well as relations between the mass and the charges. Moreover, we
simplify Masood-ul-Alam's proof as we do not require any non-trivial extensions of Witten's positive
mass theorem. We also obtain partial results on the uniqueness problem for general harmonic maps.
\end{abstract}

\newpage

\section{Introduction.}

In 1967 W. Israel proved (roughly speaking) that static, asymptotically flat
(AF) vacuum spacetimes
with non-degenerate, connected horizons are Schwarzschild \cite{WI1}. His method is based on the
integration of  two ``divergence identities''
 constructed from the norm $V$ of the static Killing vector
$\vec{\xi}$ and from the induced metric $\widehat g$ on a slice $\Sigma$ orthogonal to it (and their
derivatives). In the sequel Israel's theorem has  been generalized to include certain matter fields.
In particular,  Israel himself also proved uniqueness in the Einstein-Maxwell
(EM) case \cite{WI2}.
Later, it was realized that the proof of this theorem  could be better understood in terms of the
$SO(2,1)$-symmetry of the ``potential space'' (i.e. of the target space of the corresponding harmonic
map) \cite{DM1,WS1}. Associated with that symmetry there are conserved currents and a suitable combination
of them yields, upon integration, a functional relationship between $V$ and the electrostatic potential
$\phi$. Using these relations one can then apply the symmetry transformations on the target space,
which reduces the problem to the vacuum case.
This observation leads immediately to a further generalization, namely to Einstein-Maxwell-dilaton
(EMD) theory with ``string coupling'' ($\alpha = 1$) \cite{WS1}. The symmetry group of this theory is a
direct product of an ``electric'' and a ``magnetic'' $SO(2,1)$-part, (and the target space is a
corresponding direct sum), whence the components can be treated individually as above. This gives
uniqueness for a three-parameter family of black hole solutions found by
Gibbons \cite{GG}. In fact, arguments along these lines apply to the much
more general case in which the target space is a symmetric space
and yield uniqueness results for solutions which arise from the
Schwarzschild family by applying those symmetry transformations of the
respective theory which preserve AF \cite{BGM}.
(We remark, however, that the uniqueness results in \cite{DM1,WS1,BGM} all contain some
errors or gaps).

An alternative strategy for proving uniqueness consists, in essence, of performing conformal
rescalings on the spatial metric $\widehat g$ suitable for applying the rigidity case of
a positive mass theorem.
The ``basic
version'', appropriate for non-degenerate black holes in the vacuum case, was found in 1987 by
Bunting and Masood-ul-Alam \cite{BM}. These authors take two copies of the region exterior to the
horizon, glue them together along the bifurcation surface and rescale the metric on this compound with
a suitable (positive) function of the norm of the static Killing vector which compactifies one of the
ends smoothly. The resulting space
is shown to be complete and with vanishing Ricci scalar and vanishing mass.
Hence, the rigidity case of the positive mass theorem implies that the
rescaled metric must be flat, and the
rest follows from the field equations in a straightforward manner.

The main advantage of the method by Bunting and Masood-ul-Alam is that it
admits disconnected horizons a priori. However, as to generalizations to matter fields, they are not so
straightforward to obtain along these lines. The first part of the strategy is to find
candidates for conformal factors by taking functions which transform the family of spherically
symmetric black hole (BH) solutions whose uniqueness is conjectured to flat space, and express these
functions in terms of $V$ and the potentials. In general  there are many possibilities if matter
is present. However, (as follows from our Theorem 2),
for harmonic maps there is in fact a
unique choice of such ``factor candidates''  as functions on the target space provided that the latter
has the same dimension as the space of spherically symmetric BH. This is the case,
in particular, for EM, where these dimensions are three. (Alternatively, the magnetic or the
electric field can in advance be removed by a trivial duality transformation, which reduces the dimensions
to two).
These ``factor candidates'' are direct generalizations of the vacuum quantities, and so the same
procedure as before yields uniqueness of the non-extreme Reissner-Nordstr\"om solution \cite{WS2,AM1}.

In EMD theory, spherically symmetric BH have been studied extensively (see, e.g. \cite{GM,GH} and the 
references therein). We first note that in this situation no duality transformation is
available to remove either the electric or the magnetic field. Assuming for the moment that the latter is
absent, there is just a two-parameter family of spherically symmetric BH which,
consequently, cannot define a unique ``factor candidate'' on the three-dimensional target space. While
Masood-ul-Alam did not give a single suitable conformal factor in this situation, he made  remarkable
observations \cite{AM2} which we reformulate as follows. Firstly, for the coupling $\alpha = 1$,
and assuming a certain relation between the mass and the charges, he found two pairs of conformal factors 
${}^{\Phi}\Omega_{\pm}$ and
${}^{\Psi}\Omega_{\pm}$ such that the Ricci scalars ${}^{\Phi}{\cal R}$ and ${}^{\Psi}{\cal R}$
corresponding to the metrics ${}^{\Phi}g^{\pm} = {}^{\Phi}\Omega_{\pm}^2 V^2 \widehat g$
and ${}^{\Psi}g^{\pm} = {}^{\Psi}\Omega_{\pm}^2 V^2 \widehat g$ satisfy
${}^{\Phi} \Omega_{\pm}^2 {}^{\Phi}{\cal R} + {}^{\Psi} \Omega_{\pm}^2 {}^{\Psi}{\cal R} \ge 0$.
Secondly, he observed that the rigidity case of Witten's  positive mass theorem has a
generalization which requires just the condition above (rather than non-negativity of each
Ricci-scalar) to give flatness of ${}^{\Phi}g^{\pm}$ and ${}^{\Psi}g^{\pm}$ provided that the masses
of these metrics also vanish. (For the general formulation of this ``conformal positive mass theorem''
 c.f. Simon \cite{WS3}). By adapting  the remaining procedure from the vacuum case, Masood-ul-Alam
then obtained uniqueness of the two-parameter subfamily of the Gibbons solutions
mentioned above \cite{AM2}.

The achievements of the present paper are threefold. Firstly, we show (in
Lemma 4) that the seemingly subtle
conformal positive mass theorem
of Masood-ul-Alam and Simon have in fact trivial proofs, based on the following fact: If the
Ricci scalars $R$ and $R'$ of two metrics $h$ and $h'$ related by a conformal rescaling
$h' = \Omega^2 h$ satisfy $R +  \Omega^2 R' \ge 0$, then the Ricci scalar
$\widetilde R$ of the metric $\widetilde h = \Omega h$ is (manifestly) non-negative (by virtue
of the standard formula for conformal rescalings). In particular, Masood-ul-Alam's uniqueness result
can be obtained by  applying this observation to $h = {}^{\Phi}g^{\pm}$ and
$h' = {}^{\Psi}g^{\pm}$, and by using the rigidity case of the standard positive mass theorem
for the metric $\widetilde h = {}^{\Phi}\Omega~ {}^{\Psi}\Omega~ V^2 \widehat g$.

Secondly, we extend (in Theorem 1) Masood-ul-Alam's uniqueness results in
EMD theory to the cases with non-vanishing magnetic field (still for the coupling $\alpha =1$) on the one
hand, and to arbitrary $\alpha$ but either vanishing magnetic field or vanishing electric field on the
other hand (while the generic case is still open). 
To obtain this result we have not only to assume that the horizon is non-degenerate,
but in addition that the mass and the charges do not satisfy the relation characterizing the
spherically symmetric BH with degenerate horizons.
As to the situation with none of the fields vanishing, it is  ``underdetermined''
in the sense that we have a four-dimensional target space with just a three-parameter family of spherically
symmetric solutions. However, as mentioned above in connection with Israel's method, this target space
splits into a direct sum on which there act ``electric'' and  ``magnetic'' $SO(2,1)$ groups,
respectively. On each component we can now define  pairs of conformal factors
${}^{\Phi}\Omega_{\pm}$  and ${}^{\Psi}\Omega_{\pm}$ in a natural manner. Thus, exploiting the group
structure in this way again reduces the problem, in essence, to the EM case.

The case of arbitrary $\alpha$ but without magnetic or electric field is the more subtle one.
We have now a three-dimensional target space, with invariance group $SO(2,1) \times SO(1,1)$,
and a two-parameter family of spherically symmetric BH found by Gibbons and Maeda
\cite{GM}. Along with the two components of the group there come again naturally two pairs of
conformal factors ${}^{\Phi}\Omega_{\pm}$ and
${}^{\Psi}\Omega_{\pm}$ such that (in the case with vanishing magnetic field) the corresponding
Ricci scalars satisfy
${}^{\Phi} \Omega_{\pm}^2 {}^{\Phi}{\cal R} + \alpha^2 \, {}^{\Psi} \Omega_{\pm}^2 {}^{\Psi}{\cal R} \ge 0$.
Now we use the following extension of the previous observation: If the Ricci scalars $R$ and $R'$
of two metrics $h$ and $h'$ related by a conformal rescaling $h' = \Omega^2 h$ satisfy
$R +  \beta \Omega^2 R' \ge 0$ for some constant $\beta$, then the Ricci scalar $\widetilde R$ of the
metric $\widetilde h = \Omega^{2\beta/(1 + \beta)} h$ is (manifestly) non-negative. Thus the
uniqueness proof  can now be completed by
taking  $\beta = \alpha^2$, $h = {}^{\Phi} g$,
$h' = {}^{\Psi} g$ and by applying the standard positive mass theorem to
$\widetilde h =  {}^{\Phi} \Omega^{2/(1+\beta)}~ {}^{\Psi} \Omega^{2 \beta/(1 + \beta)} V^2\widehat g$.

Thirdly, we consider ``coupled harmonic maps''  in general. 
In Theorem 2 we show that ``candidates''  for conformal factors are determined uniquely on a 
subset of the target space corresponding to spherically symmetric BH. 
In order to obtain a uniqueness proof these ``candidates'' would have to
(i) be extended suitably
to the whole target space if the spherically symmetric BH subset is smaller than the whole target space
and (ii) be 
shown to be positive, having the right behaviour at infinity and at the horizon for 
every coupled harmonic map (without the assumption of spherical symmetry) 
and with rescaled Ricci scalar being non-negative. 
We discuss Theorem 2 with the EMD case serving as an example. 

We finally recall that, in the vacuum case P. Chru\'sciel was able to
extend the uniqueness proof such that horizons with degenerate components \cite{PC1}
are admitted a priori, and he also obtained a certain uniqueness result for degenerate horizons
in the presence of electromagnetic fields \cite{PC2}. The idea is to use an alternative conformal
rescaling due to Ruback \cite{PR} (which avoids compactification) and a suitably generalized positive
mass theorem by Bartnik and Chru\'sciel \cite{BAC} which allows ``holes''. To obtain a further
generalization including dilatons with the present methods would require  a ``conformal'' version of
this positivity result, which is not known.

\section{Basic Definitions}

\begin{definition}
A smooth spacetime $\Mg$ is called a static non-degenerate black hole iff
the following conditions are satisfied.

\begin{itemize}
\item[(1.1)] $\Mg$ admits a hypersurface orthogonal Killing vector
$\vec{\xi}$
(i.e. $\xi_{\left[\alpha\right . }\nabla_{\beta}\xi_{\left . \gamma\right]}=0$)
with a non-degenerate Killing horizon $\HH$.
\item[(1.2)] The horizon $\HH$ is of bifurcate type, i.e.
the closure $\ov{\HH}$ of $\HH$ contains points where the Killing
vector $\vec{\xi}$ vanishes.
\item[(1.3)] $\Mg$ admits an asymptotically flat hypersurface
$\Sigma$ which is orthogonal to the Killing vector $\vec{\xi}$ and such
that $V^2 = - \xi^{\alpha} \xi_{\alpha}  \rightarrow 1$ at infinity and
$\partial \Sigma \subset \ov{\HH}$.
\end{itemize}
\end{definition}
{\bf Remarks.}
\begin{itemize}
\item[1.]
A Killing horizon $\HH$ is a null hypersurface where $\vec{\xi}$ is null, non-zero and tangent
to $\HH$. The surface gravity $\kappa$  of $\HH$ is defined as
$\nabla_{\alpha} V^2 |_{\HH} = 2 \kappa \xi_{\alpha} |_{\HH}$; it is necessarily constant on each
connected component of $\HH$ (see \cite{BC}) and nonzero (by definition) for non-degenerate horizons.
\item[2.]
Ra\'cz and Wald  have shown that condition (1.2) is satisfied in
most cases of interest in which (1.1) holds. More precisely, when
the Killing vector is complete with orbits diffeomorphic to
$\Bbb{R}$ and $\HH$ is a trivial bundle over the set of orbits
$\HH /\vec{\xi}$ of the Killing vector, then a non-degenerate
horizon is of bifurcate type or else the geodesics tangent to the
Killing vector $\vec{\xi}$ reach a curvature singularity for a
finite value of the affine parameter \cite{RW2}. Similarly,
condition (1.2) is automatically satisfied in stationary, globally
hyperbolic spacetimes containing no white hole region (cf.
\cite{RW2}; and see \cite{RW1} for the precise conditions). Thus,
we could replace condition (1.2) by any of these global conditions
on the spacetime.
\end{itemize}

We also remark that our only global condition is contained in (1.3).
By AF we mean the following

\begin{definition}
A Riemannian manifold $(\Sigma, h)$ is called asymptotically flat iff
\begin{itemize}
\item[(2.1)] Every ``end'' $\Sigma^{\infty}$, (which is a connected component
of $ \ov{\Sigma} \setminus \{ \mbox{a sufficiently large}$ $\mbox{compact set} \}$) is diffeomorphic
to $\Bbb{R}^3 \setminus B$, where $B$ is a closed ball.
\item[(2.2)] On $\Sigma^{\infty}$ the metric satisfies
(in the cartesian coordinates defined by the diffeomorphism above and with
$r^2 = \sum_{i} (x^{i})^2$)
\begin{eqnarray}
\label{af}
h_{ij} - \delta_{ij} = \mbox{O}^2 (r^{-\delta}) \,\,\, \mbox{for some }
\delta >0.
\end{eqnarray}
\end{itemize}
(A function $f(x^i)$ is said to be $O^{k}(r^{\alpha})$, $k \in
\Bbb{N}$, if $f(x^i) = \mbox{O} (r^{\alpha})$, $\partial_j f(x^i)
= O(r^{\alpha-1})$ and so on for all derivatives up to and including the $k$th
ones).
\end{definition}
{\bf Remarks.}
\begin{itemize}
\item[1.]
In the definition above, $\ov{\Sigma}$ is the topological closure of $\Sigma$, and
$ \widehat g$ is the
induced metric on $\Sigma$. Notice that our definition implies, in particular, that $\ov{\Sigma}$
is complete in the metric sense.

\item[2.]
Let $q$ be a fixed point of $\vec{\xi}$ on $\ov{\HH}$ (i.e. $q \in \ov{\HH}$ and $\vec{\xi}(q)=0$),
which exists by assumption (1.2). Then, the connected component of the set $\{p|~\vec{\xi}(p)=0 \}$
containing $q$ is a smooth, embedded, spacelike, two-dimensional submanifold of $\M$
\cite{RBo,PC1}.
Such a component is called a bifurcation surface. By assumption
(1.3), any connected component $(\partial \Sigma)_{\alpha}$ of the topological boundary of $\Sigma$
is contained in the closure of the Killing horizon. Thus, (section 5 in \cite{RW2}),
$(\partial \Sigma)_{\alpha}$ must be a subset of one of the bifurcation surfaces of $\vec{\xi}$.
Furthermore, the induced metric $\widehat g$ on the hypersurface $\Sigma$
can be smoothly extended to
$\Sigma \cup (\partial \Sigma)_{\alpha}$  (see Proposition 3.3 in
\cite{PC1}). Hence
$(\ov{\Sigma},\widehat g)$  is a smooth Riemannian manifold with boundary.
\end{itemize}

Next we define the concept of ``coupled harmonic map''  and 
``massless coupled harmonic map'' between manifolds.

\begin{definition}
\label{coupled}
A coupled harmonic map is a $C^2$ map $\Upsilon : \Sigma \rightarrow \V$ between the
manifolds $(\Sigma,g)$ and $(\V,\gamma)$, (with $g$ a positive definite metric and $\gamma$
any metric), which extremizes the Lagrangian (-density)
\begin{eqnarray}
\label{chmlag}
L = \sqrt{det \, g}
\left [ R - \gamma_{ab} (\Upsilon(x)) g^{ij} \nabla_i \Upsilon^a(x) \nabla_j \Upsilon^b(x) \right ],
\end{eqnarray}
upon independent variations with respect to $g_{ij}$ and
$\Upsilon^a(x)$ (Here $\nabla$ is the covariant derivative and $R$
is the Ricci scalar with respect to $g$, and $\Upsilon^c(x)$ is
the expression of $\Upsilon$ in local coordinates of $\V$). The
Euler-Lagrange equations of (\ref{chmlag}) are called coupled
harmonic map equations and read explicitly
\begin{eqnarray}
\label{chm1}
\nabla_i \nabla^i \Upsilon^a(x) + \Gamma^{a}_{bc}\left(\Upsilon(x)
\right) \nabla_i \Upsilon^b(x) \nabla^i \Upsilon^c(x) = 0, \\
\label{chm2}
R_{ij}(x) = \gamma_{ab} (\Upsilon(x)) \nabla_i \Upsilon^a(x)
\nabla_j \Upsilon^b(x).
\end{eqnarray}
where $R_{ij}$ is the Ricci tensor of $g$ and $\Gamma^a_{bc}$ are
the Christoffel symbols of the metric $\gamma$.
\end{definition}
{\bf Remarks.}
\begin{enumerate}
\item
The definition above generalizes the notion of ``harmonic map'' which
has as Lagrangian only the second term in (\ref{chmlag}), with prescribed metric
$g$ and with $\Upsilon^a(x)$ as dynamical variable.

\item Below we will consider ``massless coupled harmonic maps'' which we define to be 
coupled harmonic maps such that 
$(\Sigma,g)$ is AF with vanishing mass, i.e. $\delta > 1$ in (\ref{af}).

\item
Coupled harmonic maps as defined above for 3-dimensional configuration spaces arise via ``dimensional
reduction'' 
from a large class of matter models (in particular for ``massless'' fields) in a spacetime 
with Killing vector \cite{BGM}. To obtain these coupled harmonic maps one can 
take as the domain manifold $\Sigma$ the space of orbits, provided this space is a
manifold.  In the static case we are dealing with, $\Sigma$ can be
envisioned as a hypersurface orthogonal to the Killing field $\vec{\xi}$ and $g =
V^2 \widehat g$, where $\widehat g$ is the induced metric on $\Sigma$.
The mass of $g$ is $M_g = M_k - M$ where $M$ is the ADM mass and $M_k$ is the ``Komar mass'' 
\cite{RBe1,PC3}. 
One can show that $M_k = M$  under rather general assumptions.
In particular, this holds if $(\Sigma, \widehat g)$ is AF and if the energy-momentum tensor
satisfies $T_{\mu\nu} = r^{-3-\epsilon}$ (see \cite{PC3};
it also follows by a slight modification of the vacuum case
\cite{RBe2}. Compare also \cite{RBe1} which is valid for complete slices). 
 In the EMD case we will show below
that asymptotic flatness alone as introduced in Definition 2 
(i.e. without the falloff requirement on $T_{\mu\nu}$) yields $M_k = M$ and hence a massless 
coupled harmonic map. In Sect. 5  we will consider massless coupled harmonic maps in general.
\end{enumerate}

\section{Einstein-Maxwell-dilaton fields}

The EMD theory is defined by the following Lagrangian (-den\-si\-ty) on $\M$
\begin{equation}
\label{Lag}
L  = \sqrt{-{}\mbox{det} \, ^4 g}({}^{4}R -
2 \nabla_{\alpha}\tau \nabla^{\alpha}\tau -
e^{- 2\alpha \tau}  F_{\alpha \beta}F^{\alpha \beta}).
\end{equation}
Here, ${}^4 g$ is a Lorentzian metric on $\M$,  $\tau$ is a scalar field, $\alpha$ is a real 
and positive (``coupling''-) constant, and the 2-form $F_{\mu\nu}$ is closed.
By the latter property (which is one of Maxwell's equations) there exists locally a vector 
potential $A_{\mu}$ such that $F_{\alpha \beta} = 2\nabla_{[\alpha}A_{\beta]}$.
Taking $g_{\mu\nu}$, $\tau$ and $A_{\mu}$ as dynamical variables in (\ref{Lag}),
variation with respect to $A_{\mu}$  implies that the 2-form
$ e^{-2\alpha \tau} \ast F_{\mu\nu} =
\frac{1}{2} e^{-2\alpha \tau} \epsilon_{\mu\nu}^{~~~\alpha\beta}F_{\alpha\beta}$ (where
$\epsilon_{\mu\nu\alpha\beta}$ is the volume form corresponding to ${}^4 g$) is also
closed (the second Maxwell equation). Hence, locally there also exists a vector potential
$C_{\mu}$ such that $\ast F_{\mu\nu} =  2e^{2\alpha \tau} \nabla_{[\mu}C_{\nu]}$.
(Alternatively, we could have taken $C_{\mu}$ as dynamical variable and
derived the existence of $A_{\mu}$).
EM is contained as the particular case
$\tau = \mbox{const}$. Other important subcases are $\alpha = 1$, which arises in string theory
and as the bosonic sector of $n=4$ supergravity, and $\alpha = \sqrt{3}$ which corresponds to
Kaluza-Klein theory (i.e. a Ricci-flat Lorentzian metric on a 5-dimensional manifold admitting a
spacelike Killing vector with certain specific properties).

We assume that on ${\cal M}$ there is a timelike, hypersurface-orthogonal Killing field $\vec{\xi}$
which also leaves the scalar and electromagnetic fields invariant. In other words, the twist vector
defined by $\omega_{\mu} = \epsilon_{\mu\nu\sigma\tau}\xi^{\nu}\nabla^{\sigma}\xi^{\tau}$ vanishes,
and we have ${\cal L}_{\xi} \tau = {\cal L}_{\xi}F_{\mu\nu}= 0$ where ${\cal L}_{\xi}$ is the Lie
derivative along $\vec{\xi}$. We further
define the electric and magnetic fields by $E_{\mu} = F_{\mu\nu}\xi^{\nu}$ and
$B_{\mu} = e^{-2 \alpha \tau} \ast F_{\mu\nu}\xi^{\nu}$. Using these definitions together with
$\omega_{\mu} = 0$ and with the Ricci identitities and the Einstein
equations, we obtain
\begin{eqnarray}
0 = \nabla_{[\mu}\omega_{\nu]} =
\epsilon_{\mu\nu\sigma\tau}R^{\sigma}_{~~\rho}\xi^{\rho}\xi^{\tau} =
2 E_{[\mu}B_{\nu]},
\label{constraint}
\end{eqnarray}
and therefore either $B_{\mu} = 0$ or $E_{\mu} = a B_{\mu}$ for some function $a$. In the
EM case ($\tau = \mbox{const.}$), it is easy to see that $a = \mbox{const.}$,
but this need not hold when the dilaton field is present.
Maxwell's and Killing's equations now imply that $\nabla_{[\mu}E_{\nu]} = 0$ and $\nabla_{[\mu}B_{\nu]} =
0$. Hence, assuming that the manifold ${\cal M}$ is simply connected, there exist
(globally) electric and magnetic potentials $\phi$ and $\psi$ defined (up to constants) by
$E_{\mu} = \nabla_{\mu}\phi$ and by $B_{\mu} = \nabla_{\mu}\psi$.
We remark that, on domains where the vectors $A_{\mu}$ and $C_{\mu}$ are
defined, we can achieve that  ${\cal L}_{\xi} A_{\mu} = {\cal L}_{\xi} C_{\mu} =
{\cal L}_{\xi}\phi = {\cal L}_{\xi}\psi = 0$ by a suitable choice of gauge
(i.e. by adding gradients of suitable functions to $A_{\mu}$ and $C_{\mu}$).
In this gauge the scalar potentials also satisfy $\phi = A_{\mu}\xi^{\mu}$ and $\psi= C_{\nu}\xi^{\nu}$.

We now write the EMD field equations as equations on a hypersurface
$(\Sigma,\widehat g)$ orthogonal to $\vec{\xi}$. In terms of the variables introduced above
they read explicitly (with $\widehat \nabla$ denoting the covariant derivative
and $\widehat \Delta$ the Laplacian with respect to $\widehat g$),

\begin{eqnarray}
\label{DelV}
\lefteqn{\widehat \Delta V  = 
V^{-1} e^{-2 \alpha \tau} \widehat\nabla_{i}\phi \widehat\nabla^{i}\phi +
V^{-1} e^{2 \alpha \tau} \widehat \nabla_{i}\psi \widehat\nabla^{i}\psi,} \\
\label{Deltau}
\lefteqn{\widehat \Delta \tau  =  - V^{-1}\widehat\nabla_{i}\tau \widehat\nabla^{i} V +
\alpha V^{-2} e^{-2 \alpha \tau} \widehat\nabla_{i}\phi \widehat\nabla^{i}\phi -
\alpha V^{-2} e^{2 \alpha \tau} \widehat\nabla_{i}\psi \widehat\nabla^{i}\psi,} \\
\label{Delphi}
\lefteqn{\widehat \Delta \phi  = 
V^{-1}\widehat\nabla_{i}V \widehat\nabla^{i}\phi -
2  \alpha \widehat\nabla_{i}\tau \widehat\nabla^{i}\phi,} \\
\label{Delpsi}
\lefteqn{\widehat \Delta \psi  = 
V^{-1}\widehat\nabla_{i}V \widehat\nabla^{i}\psi +
2 \alpha \widehat\nabla_{i}\tau \widehat\nabla^{i}\psi,}\\
\label{Riccg}
\lefteqn{\widehat R_{ij}  =
 V^{-1}\widehat\nabla_{i}\widehat\nabla_{j}V + 2 \widehat\nabla_{i}\tau \widehat\nabla_{j}\tau -
{}} \nonumber\\
& & - V^{-2} e^{-2 \alpha \tau} (2\widehat\nabla_{i}\phi \widehat\nabla_{j}\phi -
\widehat g_{ij}\widehat\nabla_{k}\phi \widehat\nabla^{k}\phi) -
V^{-2} e^{2 \alpha \tau} (2\widehat\nabla_{i}\psi \widehat\nabla_{j}\psi -
\widehat g_{ij}\widehat\nabla_{k}\psi \widehat\nabla^{k}\psi),
\end{eqnarray}
where $\widehat R_{ij}$ is the Ricci tensor of $\widehat g$.
For the trace of (\ref{Riccg}) we obtain
\begin{equation}
\label{Rg}
\widehat R  = 2 \widehat\nabla_{i}\tau \widehat\nabla^{i}\tau +
2 V^{-2} e^{-2 \alpha \tau} \widehat\nabla_{i}\phi\widehat\nabla^{i}\phi +
2 V^{-2} e^{2 \alpha \tau} \widehat\nabla_{i}\psi\widehat\nabla^{i}\psi.
\end{equation}
We first give a lemma on the behaviour of the fields on the horizon.
\begin{lemma}
For static Einstein-Maxwell-dilaton non-degenerate black holes, there hold
the following relations on the boundary $\partial \Sigma$
\begin{eqnarray}
\label{lim}
\left .  \na^i  V \na_i \tau \right |_{\partial \Sigma} = 0,
\hspace{1cm}
\left .  \na^i  V \na_i \phi \right |_{\partial \Sigma} = 0,
\hspace{1cm}
\left . \na^i  V \na_i \psi \right |_{\partial \Sigma} = 0.
\label{condboundary}
\end{eqnarray}
\end{lemma}

{\it Proof.}~Recall that the induced metric $\widehat g$ on the
hypersurface $\Sigma$ can be smoothly extended to $\Sigma \cup
(\partial \Sigma)_{\alpha}$  (see Proposition 3.3 in \cite{PC1}).
Since $\tau$ was assumed to be $C^2$, it follows from (\ref{Rg})
that $ V^{-2} e^{-2 \alpha \tau} \widehat\nabla_{i}\phi\widehat\nabla^{i}\phi $
and $ V^{-2} e^{2 \alpha \tau} \widehat\nabla_{i}\psi\widehat\nabla^{i}\psi$
have regular extensions to $\partial \Sigma$. Now the first
equation in (\ref{lim}) follows from (\ref{Deltau}) while the
remaining two equations follow from (\ref{Delphi}) and
(\ref{Delpsi}). $\hfill \Box$

\medskip

We can bring equations (\ref{DelV})-(\ref{Riccg}) to the form of a
coupled harmonic map between $(\Sigma, V^2 \widehat g)$ and the
four-dimensional target manifold $\V$ defined by $(V, \tau,
\phi,\psi) \in \Bbb{R}^{+} \times \Bbb{R} \times \Bbb{R} \times
\Bbb{R}$ endowed with the metric
\begin{eqnarray}
ds^2 = \gamma_{ab} dx^a dx^b =
2 V^{-2} dV^2 + 2 {d \tau ^2}
- 2 V^{-2} ( e^{-2\alpha \tau} d\phi^2 +  e^{2 \alpha \tau} d \psi^2).
\label{targetEMD}
\end{eqnarray}

For reasons discussed in Sect. 5, our results on this model will be restricted to 
three special cases, namely $\psi = 0$, $\alpha = 1$ and  $\phi = 0$. In each
case, it is useful for our purposes to  parametrize the target space
$\V$ by variables (denoted by $\Phi_{A}$ and $\Psi_{A}$ (A = -1,0,1))
in terms of which the isometry group of $(\V,\gamma)$ acts linearly.
The definitions of $\Phi_{A}$ and $\Psi_{A}$ are different in the three
cases, but we treat these cases independently and therefore use below
the same symbols for simplicity.

Thus, in terms of the auxiliary variables
$\gamma_{\beta} = V e^{\beta \tau}$, $\beta \in \Bbb{R}$,
$\widetilde \phi = \sqrt{\alpha^2 + 1}~\phi $ and
$\widetilde \psi = \sqrt{\alpha^2 + 1}~\psi $ we define

\bigskip

$\psi = 0$:

\parbox{8cm}
{\begin{eqnarray*}
\Phi_{-1} &  = & \frac{1}{2}[\gamma_{\alpha} - \gamma_{\alpha}^{-1}(\widetilde\phi^2 + 1)],\\
\Phi_{0} & = & \gamma_{\alpha}^{-1}\widetilde\phi,\\
\Phi_{1} &  = & \frac{1}{2}[\gamma_{\alpha} - \gamma_{\alpha}^{-1}(\widetilde\phi^2 - 1)],
\end{eqnarray*}}
\parbox{5cm}
{\begin{eqnarray*}
\Psi_{-1}& = & \frac{1}{2}(\gamma_{-1/\alpha} - \gamma^{-1}_{-1/\alpha}),\\
\Psi_{0} &  = & 0, \\
\Psi_{1}& = & \frac{1}{2}(\gamma_{-1/\alpha} +  \gamma^{-1}_{-1/\alpha}).
\end{eqnarray*}}

\medskip

$\alpha = 1$:

\parbox{8cm}
{\begin{eqnarray*}
\Phi_{-1} &  = & \frac{1}{2}[\gamma_1 - \gamma_1^{-1}(\widetilde \phi^2 + 1)],\\
\Phi_{0} & = & \gamma_{1}^{-1}\widetilde \phi,\\
\Phi_{1} &  = & \frac{1}{2}[\gamma_1 - \gamma_1^{-1}(\widetilde \phi^2 - 1)],
\end{eqnarray*}}
\parbox{6cm}
{\begin{eqnarray*}
\Psi_{-1} &  = & \frac{1}{2}[\gamma_{-1} - \gamma_{-1}^{-1}(\widetilde\psi^2 + 1)],\\
\Psi_{0} & = & \gamma_{-1}^{-1} \widetilde \psi,\\
\Psi_{1} &  = & \frac{1}{2}[\gamma_{-1} - \gamma_{-1}^{-1}(\widetilde \psi^2 - 1)].
\end{eqnarray*}}

\medskip

$\phi = 0$:

\parbox{8cm}
{\begin{eqnarray*}
\Phi_{-1}& = & \frac{1}{2}(\gamma_{1/\alpha} - \gamma^{-1}_{1/\alpha}),\\
\Phi_{0} &  = & 0, \\
\Phi_{1}& = & \frac{1}{2}(\gamma_{1/\alpha} +  \gamma^{-1}_{1/\alpha}),
\end{eqnarray*}}
\parbox{6cm}
{\begin{eqnarray*}
\Psi_{-1} &  = & \frac{1}{2}[\gamma_{-\alpha} -
\gamma_{-\alpha}^{-1}(\widetilde\psi^2 + 1)],\\
\Psi_{0} & = & \gamma_{-\alpha}^{-1}\widetilde\psi,\\
\Psi_{1} &  = & \frac{1}{2}[\gamma_{-\alpha} -
\gamma_{-\alpha}^{-1}(\widetilde\psi^2 - 1)].
\end{eqnarray*}}

Capital indices are raised and lowered  with the metric $\eta_{AB} = \mbox{diag}(1,-1,-1)$.
Since we define here six variables out of the four ones $V, \tau, \phi$ and
$\psi$ there must be two constraints, which read
$\Phi_{A} \Phi^{A} = -1 = \Psi_{B}\Psi^{B}$. We also introduce
the following quantities (which are in general {\it not} Ricci tensors of any metric)

\parbox{6cm}
{\begin{eqnarray*}
{}^{\Phi}R_{ij} &  = & \nabla_{i}\Phi^{A} \nabla_{j}\Phi_{A}, \\
{}^{\Phi}R & = & g^{ij} \, {}^{\Phi}R_{ij} ,
\end{eqnarray*}}
\parbox{6cm}
{\begin{eqnarray*}
{}^{\Psi}R_{ij} & = & \nabla_{i}\Psi^{A} \nabla_{j}\Psi_{A},\\
{}^{\Psi}R & = & g^{ij} \, {}^{\Psi}R_{ij},
\end{eqnarray*}}

\noindent where $\nabla$ denotes the covariant derivative of $g = V^2
\widehat g$.
We now write the coupled harmonic map field equations in terms of these
variables.
Since here and henceforth the case $\phi = 0$ arises from the case $\psi = 0$ via the exchange
$\Phi \leftrightarrow \Psi$, we only give the latter case explicitly.
($\Delta$ denotes the Laplacian with respect to $g$).

\begin{eqnarray}
\label{DelPhiPsi}
\Delta \Phi_{A} & = & {}^{\Phi}R \, \Phi_{A}, \qquad
\Delta \Psi_{A} = {}^{\Psi}R \, \Psi_{A}, \\
\label{Riccpsi}
\psi = 0: \qquad
R_{ij} & = & \frac{2}{1 + \alpha^2}
({}^{\Phi}R_{ij} + \alpha^2~{}^{\Psi}R_{ij}), \\
\label{Riccal}
\alpha = 1: \qquad
R_{ij} & = & {}^{\Phi}R_{ij} + {}^{\Psi}R_{ij}.
\end{eqnarray}
These equations can be obtained by varying the Lagrangian
(-densities)

\begin{eqnarray*}
\psi & = & 0: \qquad L = \sqrt{- \mbox{det} \, g} \left ( \frac{1 + \alpha^2}{2} R +
\frac{g^{ij}\nabla_{i}\Phi^{A} \nabla_{j}\Phi_{A}}{\Phi_{A}\Phi^{A}} +
\alpha^2\frac{g^{ij}\nabla_{i}\Psi^{A} \nabla_{j}\Psi_{A}}{\Psi_{A}\Psi^{A}}
\right ),\\
\alpha & = & 1: \qquad L = \sqrt{- \mbox{det} \, g} \left ( R +
\frac{g^{ij}\nabla_{i}\Phi^{A} \nabla_{j}\Phi_{A}}{\Phi_{A}\Phi^{A}} +
\frac{g^{ij}\nabla_{i}\Psi^{A} \nabla_{j}\Psi_{A}}{\Psi_{A}\Psi^{A}} \right ).
%
\end{eqnarray*}
independently with respect to $g_{ij}$, $\Phi_{A}$ and $\Psi_{A}$ and imposing
(afterwards) the constraints $\Phi_{A} \Phi^{A} = -1 = \Psi_{B}\Psi^{B}$.

The Lagrangian, the constraints and the field equations are invariant
under the isometry group of the target space metric
(\ref{targetEMD}). In terms of the variables $\Phi^A$ and $\Psi^A$ the
invariance transformations take the simple form
\begin{eqnarray*}
\Phi_{A}' = {}^{\Phi}L_{~~B}^{A} \Phi^{B}, \qquad
\Psi_{A}' = {}^{\Psi}L_{~~B}^{A} \Psi^{B},
\end{eqnarray*}
where ${}^{\Phi}L_{~~B}^{A}$ and ${}^{\Psi}L_{~~B}^{A}$
satisfy $\eta_{AB}L_{~~~C}^{\dagger A}L_{~~D}^{B}= \eta_{CD}$,
i.e. they are elements of $SO(2,1)$.
Since for vanishing magnetic and electric fields we have
$\Psi_{0} = 0$ and $\Phi_{0} = 0$, respectively, the corresponding
symmetry groups are $SO(2,1) \times SO(1,1)$, while in the case
$\alpha = 1$ we have the full group $SO(2,1) \times SO(2,1)$.
We will also use the notation
$\Phi_{a} = \{\Phi_{-1},\Phi_{0}\}$,  $\Psi_{a} = \{\Psi_{-1},\Psi_{0}\}$,
and move these indices with the metric $\eta_{ab} = \mbox{diag}(1,-1)$.

In the case $\alpha = 1$ Gibbons has given the general
(three-parameter-) family of spherically symmetric solutions \cite{GG}
(which we write here in harmonic coordinates)

\begin{eqnarray}
\label{GGPhiPsi}
\Phi_{a} & = &  \frac{{}^{\Phi}M_{a}}{\sqrt{\rho^{2} - A^{2}}}, \qquad
\Psi_{a} =  \frac{{}^{\Psi}M_{a}}{\sqrt{\rho^{2} - A^{2}}},\\
\label{GGg}
ds^{2} & = & d\rho^{2} + (\rho^{2} - A^{2})
(d\theta^{2} + \sin^{2}\theta d\phi^{2}), 
\end{eqnarray}
where ${}^{\Phi}M_{a}$, ${}^{\Psi}M_{a}$ and $A \ge 0$ are constants satisfying
${}^{\Phi}M_{a}{}^{\Phi}M^{a} = {}^{\Psi}M_{a}{}^{\Psi}M^{a} = A^2$.
(By sub- and superscripts on multipole moments we always mean indices and not
exponents).

The spherically symmetric solutions in the other two cases ($\psi = 0$ and $\phi = 0$;
c.f. Gibbons and Maeda \cite{GM}) are given by the two-parameter
subfamilies with $\Psi_{0} = 0$ and $\Phi_{0} = 0$, respectively.
Clearly, the horizon is located at $\rho = A$, with $A = 0$
characterizing the degenerate case.

In the next section we will prove uniqueness of these classes of
solutions (in the non-degenerate case).

\medskip

We now examine in detail the asymptotic structure of the fields introduced
above. The complete analysis is somewhat involved, but consists in essence
of assembling and adapting bits and pieces available in the literature.
The whole procedure is also quite similar to the ``instanton'' case considered in \cite{MS}.

\begin{lemma}
\label{lem2}
 On an end $(\Sigma^{\infty}, g)$ of a static,
asymptotically flat solution of (\ref{DelPhiPsi})-(\ref{Riccal})
there is a coordinate system $x^{i}$ (in general different from
the one of Def.~1 but still called $x^{i}$) and there exist
constants ${}^{\Phi}M^{a}$, ${}^{\Phi}M^{a}_{i}$ ${}^{\Psi}M^{a}$
and ${}^{\Psi}M^{a}_{i}$ such that

\begin{eqnarray}
\label{expPhi}
\Phi^{a} & = & \frac{{}^{\Phi}M^{a}}{r} +
\frac{{}^{\Phi}M^{a}_{i}x^{i}}{r^{3}}
+ O^{\infty}(\frac{1}{r^{3}}), \\
\label{expPsi}
\Psi^{a} & = & \frac{{}^{\Psi}M^{a}}{r} +
\frac{{}^{\Psi}M^{a}_{i}x^{i}}{r^{3}}
+ O^{\infty}(\frac{1}{r^{3}}), \\
\label{expg}
g_{ij} & = & \delta_{ij} +
\frac{{}^{\Phi}M^{a}~{}^{\Phi}M_{a} + {}^{\Psi}M^{a}~ {}^{\Psi}M_{a}}{r^{4}}
(\delta_{ij}r^{2} - x^{i}x^{j})+ O^{\infty}(\frac{1}{r^{3}}).
\end{eqnarray}
\end{lemma}
{\it Proof.}
The definition of asymptotic flatness (\ref{af}) implies that
$\widehat R = O(r^{-2-\delta})$ for some $\delta > 0$ and hence
(by adjusting constants suitably) we have, from (\ref{Rg}),
\begin{eqnarray*}
\tau = O^1(r^{-\epsilon}), \qquad \phi = O^1(r^{-\epsilon}), \qquad \psi= O^1(r^{-\epsilon}),
\qquad \mbox{for some}~~~\epsilon > 0.
\end{eqnarray*}
Using next the full equation (\ref{Riccg}) we obtain
$\widehat \nabla_{i}\widehat \nabla_{j}V = O(r^{-2-\epsilon})$. To get information
on $V$ and its partial derivatives, namely
\begin{eqnarray*}
1 - V = O^2(r^{-\epsilon}),
\end{eqnarray*}
requires an iterative procedure which we take over from Proposition 2.2 of \cite{BEC}
(compare also lemma 5 of \cite{MS}).
Standard results on the inversion of the Laplacians in (\ref{DelV})-(\ref{Delpsi})
(Corollary 1 of Theorem 1 in \cite{NM}) then yield
\begin{eqnarray*}
\tau = O^2(r^{-\epsilon}), \qquad
\phi = O^2(r^{-\epsilon}), \qquad \psi= O^2(r^{-\epsilon}).
\end{eqnarray*}
It is now useful to pass to the variables $g_{ij}$, $\Phi_{a}$ and $\Psi_{a}$
which have the asymptotic behaviour
\begin{equation}
\label{FallO2}
\Phi_{a} = O^2(r^{-\epsilon}), \qquad \Psi_{a} = O^2(r^{-\epsilon}),
\qquad g_{ij} = \delta_{ij} + O^2(r^{-\epsilon})
\end{equation}
and to introduce harmonic coordinates, which preserves these falloff properties. Then we can
write (\ref{DelPhiPsi})-(\ref{Riccal}) in the form (the subsequent step follows an idea of
Kennefick and \'O Murchadha \cite{KM} and has been erroneously omitted
in \cite{MS})
\begin{equation}
\label{DelO2}
g^{ij}\partial_i\partial_j \Phi_a = O(r^{-2 - 3\epsilon}), \hspace{3mm}
g^{ij}\partial_i\partial_j \Psi_a = O(r^{-2 - 3\epsilon}), \hspace{3mm}
g^{ij}\partial_i\partial_j g_{kl} = O(r^{-2 - 2\epsilon}).
\end{equation}
Inversion of the Laplacians now yields (\ref{FallO2}) but with $2\epsilon$ instead of $\epsilon$.
Iterating this procedure sufficiently many times, we can improve the falloff on the r.h. sides of
(\ref{DelO2}) to $O^2(r^{-3 - \beta})$ for some $\beta > 0$.
Following now \cite{SB} and \cite{WS4}, (but keeping here harmonic coordinates for simplicity) we can
write these equations as
\begin{eqnarray*}
\triangle \Phi_{a}  =  O(r^{-3-\beta}) \qquad
\triangle \Psi_{a}  =  O(r^{-3-\beta}) \qquad
\triangle g_{ij}  =  O(r^{-3-\beta})
\end{eqnarray*}
where $\triangle$ is now the flat Laplacian. Inversion yields the
monopole terms in (\ref{expPhi}) and (\ref{expPsi}), (while the
monopole term is absent in (\ref{expg}) due to the harmonic gauge
condition), and  remaining terms of $O^2(r^{-1-\beta})$. Finally,
the last procedure can also be iterated to give
(\ref{expPhi})-(\ref{expg}) as they stand. $\hfill \Box$ \\ \\
{\bf Remarks.}
\begin{itemize}
\item[1.]

In the coordinate system introduced above, we can write the (Komar-)mass $M$
and define a dilaton charge $D$ and electric and magnetic charges $Q$ and $P$ as
follows:
\begin{equation}
\mbox{lnV} = \frac{M}{r} + O(\frac{1}{r^2}) ~~~
\tau =  \frac{D}{r} + O(\frac{1}{r^2}) ~~~
\phi = \frac{Q}{r} + O(\frac{1}{r^2}) ~~~
\psi =  \frac{P}{r} + O(\frac{1}{r^2})
\end{equation}
The relation between these ``multipole moments'' and the quantities 
${}^{\Phi}M^{a}$ and ${}^{\Psi}M^{a}$ is

$\psi = 0$:
\parbox{8cm}
{\begin{eqnarray*}
{}^{\Phi}M_{-1} &  = &  - M + \alpha D \\
{}^{\Phi}M_{0} &  = & \sqrt{\alpha^2 + 1}~Q  
\end{eqnarray*}}
\parbox{5cm}
{\begin{eqnarray*}
{}^{\Psi}M_{-1} &  = &  -M - \alpha^{-1}D\\
{}^{\Psi}M_{0} &  = &  0 
\end{eqnarray*}}

$\alpha = 1$:
\parbox{8cm}
{\begin{eqnarray*}
{}^{\Phi}M_{-1} &  = &  - M +  D \\
{}^{\Phi}M_{0} &  = & \sqrt{2}~Q  
\end{eqnarray*}}
\parbox{5cm}
{\begin{eqnarray*}
{}^{\Psi}M_{-1} &  = &  -M - D\\
{}^{\Psi}M_{o} &  = &  \sqrt{2}~P 
\end{eqnarray*}}

and similar for $\phi = 0$.

\item[2.]
As elaborated in \cite{WS4}, the expansion (\ref{expPhi})-(\ref{expg})
can in fact be pursued to arbitrary orders to give multipole expansions of a rather simple structure.

\end{itemize}

\section{The uniqueness proof}

In the previous section we described three special cases of EMD theory whose target spaces have similar 
group structures. We have exposed the theory in a way which makes these structures manifest by choosing 
variables which linearize the group action. This allows us to perform the uniqueness proof in close 
analogy with the electromagnetic case \cite{WS2,AM1}. The analogy suggests, in particular, the following 
choice of conformal factors,
\begin{eqnarray}
\label{conffac}
{}^{\Phi}\Omega_{\pm} = \frac{1}{2}(\Phi_1 \pm 1), \qquad
{}^{\Psi}\Omega_{\pm} = \frac{1}{2}(\Psi_1 \pm 1),
\end{eqnarray}
and the rescaled metrics are denoted by
\begin{eqnarray*}
{}^{\Phi} g^{\pm}_{ij} = {}^{\Phi}\Omega_{\pm}^2 g_{ij}, \qquad
{}^{\Psi} g^{\pm}_{ij} = {}^{\Psi}\Omega_{\pm}^2 g_{ij}.
\end{eqnarray*}
We also define
${}^{\Phi}A_{\pm} = - {}^{\Phi}M^{-1} \pm |{}^{\Phi}M^{0}|$ and
${}^{\Psi}A_{\pm} = - {}^{\Psi}M^{-1} \pm |{}^{\Psi}M^{0}|$
in terms of the quantities introduced in (\ref{expPhi}) and (\ref{expPsi}),
while in terms of the charges $M,D,P$ and $Q$ (c.f. remark 1 at the end of Sect.3)
we have \\
$\psi = 0$:
\parbox{8cm}
{\begin{eqnarray}
{}^{\Phi}A_{\pm} = M - \alpha D \pm \sqrt{1 + \alpha^2} |Q| \nonumber
\end{eqnarray}}
\parbox{5cm}
{\begin{eqnarray}
\label{psiA}
{}^{\Psi}A_{\pm} = M + \alpha^{-1} D 
\end{eqnarray}}
$\alpha = 1$:
\parbox{7.5cm}
{\begin{eqnarray}
{}^{\Phi}A_{\pm} = M - D \pm \sqrt{2} |Q| \nonumber
\end{eqnarray}}
\parbox{6cm}
{\begin{eqnarray}
\label{alA}
{}^{\Psi}A_{\pm} = M + D \pm \sqrt{2} |P|
\end{eqnarray}}
 
Finally, we introduce ${}^{\Phi}A^2 =  {}^{\Phi}A_{+}{}^{\Phi}A_{-}$ and 
${}^{\Psi}A^2 = {}^{\Psi}A_{+} {}^{\Psi}A_{-}$.
In the following Lemma we show (among other things) that these quantities 
are in fact non-negative. 

\begin{lemma}
\label{lem}
Let $(\Sigma,g,\Phi_a,\Psi_a)$ be non-degenerate black hole solutions of
(\ref{DelPhiPsi})-(\ref{Riccal}) with ${}^{\Phi}A_{-} \ne 0$ and ${}^{\Psi}A_{-} \ne 0$. 
Then
\begin{enumerate}
\item
$(\ov{\Sigma},{}^{\Phi} g^{+})$ and
$(\ov{\Sigma},{}^{\Psi} g^{+})$ are asymptotically flat Riemannian
spaces with $C^{2}$- metrics and with vanishing mass.
\item
$(\ov{\Sigma},{}^{\Phi} g^{-})$
$(\ov{\Sigma},{}^{\Psi} g^{-})$ admit one-point compactifications
$\widetilde{\Sigma}=\ov{\Sigma} \cup \Gamma$ such that
$(\widetilde{\Sigma},{}^{\Phi} g^{-})$
$(\widetilde{\Sigma},{}^{\Psi} g^{-})$ are complete Riemannian spaces
with $C^2$- metrics.
\item
The spaces  $(\ov{\Sigma},{}^{\Phi} g^{+})$ and $(\ov{\Sigma},{}^{\Psi} g^{+})$
 can be glued together  with $(\ov{\Sigma},{}^{\Phi} g^{-})$ and
$(\ov{\Sigma},{}^{\Psi} g^{+})$, respectively, to give Riemannian spaces
$({\cal N}, {}^{\Phi} g)$ and $({\cal N}, {}^{\Psi} g)$ with
$C^{1,1}-metrics$.
\end{enumerate}
\end{lemma}
{\it Proof.}
The proof is identical in all three cases discussed in the preceding section
(the coupling constant $\alpha$ does not appear).
Moreover, since the proof consists of the identical ``$\Phi$''- and
``$\Psi$''-parts we only give the former explicitly.

We first show that  ${}^{\Phi}\Omega_{\pm} > 0$.
We define the quantities
${}^{\Phi}\Xi_{\pm} = (1 \pm \Phi_{0})(\Phi_{1} - \Phi_{-1})^{-1} - 1$
which satisfy ${}^{\Phi}\Xi_{+}~{}^{\Phi}\Xi_{-} =
4 {}^{\Phi}\Omega_{-}(\Phi_{1} - \Phi_{-1})^{-1}$ and we note that
 $(\Phi_{1} - \Phi_{-1}) > 0$ since $V > 0$.
Moreover, by a straightforward calculation and by using (\ref{DelPhiPsi})-(\ref{Riccal})
we find that
\begin{eqnarray}
\label{DelXi}
-\nabla^{i}[(\Phi_{1} - \Phi_{-1})^{2} \nabla_{i} {}^{\Phi}\Xi_{\pm}] =
\Delta(\Phi_{1} - \Phi_{-1}) = {}^{\Phi}R (\Phi_{1} - \Phi_{-1}) = \hspace{2cm}
\nonumber \\
= (\nabla_{i} {}^{\Phi}\Xi_{+})(\nabla^{i} {}^{\Phi}\Xi_{-})(\Phi_{1}-\Phi_{-1})^3.
\end{eqnarray}
By the maximum principle, the quantities ${}^{\Phi}\Xi_{\pm}$ take
on their extrema on the boundary, i.e. either on $\partial\Sigma$
or at infinity. Since the BH is non-degenerate, $W
\equiv \widehat \nabla_i V \widehat \nabla^{i} V$ is non-zero at
$\partial \Sigma$. Hence $\widehat n_i = - W^{-1/2}
\widehat\nabla_i V$ is a unit outward  normal to $\partial\Sigma$.
Using Lemma 1, we obtain
$\widehat n^{i}\widehat\nabla_{i} {}^{\Phi}\Xi_{\pm} < 0$
 on
$\partial\Sigma$ and so ${}^{\Phi}\Xi_{\pm}$ must in particular
take on their maxima at infinity where they approach zero, from
(\ref{expPhi}). Hence ${}^{\Phi}\Xi_{\pm} < 0$ on $\ov{\Sigma}.$
This proves the positivity of ${}^{\Phi}\Omega_{-}$, and obviously
we have ${}^{\Phi}\Omega_{+} > {}^{\Phi}\Omega_{-}$.
Observe now that the asymptotic behaviour (\ref{expPhi}) and (\ref{expPsi}) yields
${}^{\Phi}\Xi_{\pm} = r^{-1}({}^{\Phi}M^{-1} \pm {}^{\Phi}M^{0}) + O(r^{-2})$
and ${}^{\Phi}\Omega_{-} = {}^{\Phi}A^2 r^{-2} + \mbox{O}^2(r^{-3})$. 
Therefore, ${}^{\Phi}\Xi_{\pm} < 0$ implies ${}^{\Phi}A_{-} \ge 0$
and  ${}^{\Phi}\Omega_{-} > 0$ implies and  ${}^{\Phi}A^2 \ge 0$
(which justifies the definition of ${}^{\Phi}A^2$).
Together with  the definitions and the assumption of this Lemma, we obtain 
 ${}^{\Phi}A_{+} \ge {}^{\Phi}A_{-} > 0$  and  ${}^{\Phi}A^2 > 0$. Hence
${}^{\Phi}\Omega_{-}$ qualifies as conformal factor for a  $C^2$- compactification. 
Further, since ${}^{\Phi}\Omega_{+} = 1 + O^2(r^{-2})$ and since $g_{ij}$ has vanishing mass, 
the latter is also true for ${}^{\Phi} g^{\pm}_{ij}$.

To do the matching we use again standard results
(see e.g. \cite{MaSe}). We first show that the induced metric on
$\partial \Sigma$ is the same on $(\ov{\Sigma}, {}^{\Phi} g^{-})$
and on $(\ov{\Sigma}, {}^{\Phi} g^{+})$. This is the case because the
metrics are ${}^{\Phi} g^{\pm} = ({}^{\Phi}\Omega_{\pm} V)^2 (V^{-2} g ) =
( {}^{\Phi}\Omega_{\pm} V)^2 \widehat g$ and $\widehat g$ extends smoothly to
$\partial \Sigma$ (see Proposition 3.3 of \cite{PC1}) and
$V {}^{\Phi}\Omega_{\pm}$ are regular at $\partial \Sigma$.
Furthermore, the explicit expressions of ${}^{\Phi}\Omega_{\pm}$
show that $V {}^{\Phi}\Omega_{+} = V {}^{\Phi}\Omega_{-}$ at $V=0$.

The other junction condition is that the second fundamental forms of $\partial \Sigma$ with respect
to the unit outward normals of $(\ov{\Sigma}, {}^{\Phi} g^{+})$ and of
$(\ov{\Sigma}, {}^{\Phi} g^{+})$
agree apart from a sign. Under a conformal rescaling $h'_{ij}= \Omega^2 h_{ij}$, the second
fundamental of a hypersurface transforms as $K'_{AB} = \Omega K_{AB} - \vec{n} (\Omega) h_{AB}$,
where $\vec{n}$ is the unit normal vector with respect to $h_{ij}$ and $h_{AB}$ is the induced
metric on the hypersurface. We recall that the boundary $\partial \Sigma$ is totally geodesic with
respect to the metric $\widehat g_{ij}$ (i.e. $\widehat K_{AB} = 0$) \cite{WI1}. A simple
calculation using (\ref{condboundary}) now
shows that $\widehat n^{i}\widehat\nabla_{i} ( {}^{\Phi}\Omega_{\pm} V) |_{\partial \Sigma} =
\mp \kappa/2$. Thus, the two second fundamental forms differ by a sign and
so the glued Riemannian space $(\N,{}^{\Phi} g)$ is $C^{1,1}$.
$\hfill \Box$ \\ \\
{\bf Remark.} 
For the spherically symmetric
solutions (\ref{GGPhiPsi}),(\ref{GGg}), we have ${}^{\Phi}A_{-} = {}^{\Psi}A_{-}  = 0$ 
iff the horizon is degenerate. In our considerations (c.f. definition 1, and the
preceding Lemma in particular) we always exclude degenerate horizons. Hence the conditions
${}^{\Phi}A_{-} \ne 0$ and  ${}^{\Psi}A_{-} \ne 0$ in Lemma \ref{lem} seem redundant. 
In fact, for connected horizons,  ${}^{\Phi}A_{-} > 0$ and ${}^{\Psi}A_{-} > 0$ 
follow directly from the ``mass formulas'' \cite{MHeu}. 
Admitting now disconnected horizons, we report here on our partially successful efforts of 
dropping requirements ${}^{\Phi}A_{-} \ne 0$ and  ${}^{\Psi}A_{-} \ne 0$.
In terms of the mass and the charges they read (restricing ourselves  for simplicity to the 
case $\psi = 0$): 
\begin{equation}
\label{minequ1}
M - \alpha D \ne \sqrt{1 + \alpha^2}|Q| \qquad \alpha M + D \ne 0.
\end{equation}
In absence of the dilaton, i.e. in EM theory, the second condition is clearly trivial, while
the first one reduces to $M \neq |Q|$. A generalization of Witten's proof 
 yields $M \geq |Q|$ if the constraints hold (but without the assumption of
staticity) \cite{GHHP,MHer} and also gives the extreme Reissner-Nordstr\"om solution in the limiting 
case $M = |Q|$ \cite{BAC}.
In the case with dilaton, it has been shown by the  technique of \cite{GHHP} that 
\begin{equation}
\label{minequ2}
\sqrt{1 + \alpha^2} M \ge |Q|
\end{equation}
\cite{GKLTT,MR}, and we conjecture that equality holds iff the solution is the degenerate 
Gibbons-Maeda one ((\ref{GGPhiPsi}), (\ref{GGg}) with $A = 0$).
Next, as an extension of the arguments by Penrose, Sorkin and Woolgar \cite{PSW} for showing positivity
of mass, Gibbons and Wells \cite{GW} claimed that the inequality 
\begin{equation}
\label{minequ3}
M - \alpha D \ge \sqrt{1 + \alpha^2}|Q|.
\end{equation}
holds, and again it is natural to conjecture that the limiting case is
precisely the one with degenerate horizons. Now observe that, if both conjectures on the limiting cases of
(\ref{minequ2}) and (\ref{minequ3}) were true, we would obtain \emph{both} inequalities (\ref{minequ1})
in the non-degenerate case. Unfortunately, however, the positive mass claim of \cite{PSW} (and hence also
(\ref{minequ3})) has so far not been established rigorously.

We now pursue an idea (applicable to all three cases $\psi = 0$, $\phi = 0$ and $\alpha = 1$)
for showing directly that ${}^{\Phi}A_{-} \ne 0$ and ${}^{\Psi}A_{-} \ne 0$ in the static
case. To exclude the case ${}^{\Phi}A_{-} = 0$, which is equivalent to
${}^{\Phi}M^{0} = \pm {}^{\Phi}M^{-1}$, we conclude indirectly: if one of these relations held,
(\ref{expPhi}) would give, for the corresponding $\Xi_{\mp}$, the expansion
${}^{\Phi}\Xi_{\mp} = r^{-3}({}^{\Phi}M^{-1}_{i}x^{i} \mp {}^{\Phi}M^{0}_{i}x^{i})
+ O^{2}(r^{-3})$. This would, however, contradict ${}^{\Phi}\Xi_{\mp} < 0$ unless
${}^{\Phi}M^{-1}_{i}x^{i} = \pm {}^{\Phi}M^{0}_{i}x^{i}$ and thus
${}^{\Phi}\Xi_{\pm} = O^{2}(r^{-3})$.
To proceed further we now write (\ref{DelXi}) on some end $\Sigma^{\infty}$ in the form
\begin{eqnarray*}
\triangle {}^{\Phi}\Xi_{\pm} =
f^{ij} \partial_{i}\partial_{j} {}^{\Phi}\Xi_{\pm} +
k^{i} \partial_{i} {}^{\phi}\Xi_{\pm}
\end{eqnarray*}
with $\triangle$ denoting (as already in Sect. 3) the flat Laplacian, and $f^{ij}$ and $k^{i}$
are smooth
functions with falloff $O(r^{-2})$. Inverting this Laplacian we observe that the leading term in the
expansion of ${}^{\Phi}\Xi_{\pm}$ must be a homogeneous solution of order $O(r^{-3})$  which, on the
other hand, must again be absent due to ${}^{\Phi}\Xi_{\mp} < 0$. 
The idea is now to iterate this procedure and  arrive at 
${}^{\Phi}\Xi_{\mp} \equiv 0$ on the end $\Sigma^{\infty}$, which would obviously contradict
${}^{\Phi} \Xi_{\mp}<0$ and show that ${}^{\Phi}A_{-} > 0$ as claimed. 
This argument would be rigorous if there were  an analytic compactification of $\Sigma$
near spatial infinity.
Known proofs of analyticity of static solutions are based on deriving regular
elliptic systems for the conformal field equations, and have been carried out for vacuum and 
for EM fields \cite{BS,WS2}. However, to simplify the algebraic complications in the latter 
case, the particular conformal factor $(\ref{conffac})$ has been employed
which requires $M > |Q|$. (In the purely asymptotic regime, this has to be 
imposed as an extra condition). It is likely that in terms of a different conformal factor
(and after substantial algebraic work) one would obtain analyticity without this requirement. 
Extensions to the EMD case should then be straightforward as well.

Our final lemma is the ``conformal positive mass'' one, whose rigidity case will be
employed later.

\begin{lemma}
\label{rescaling}
Let $({\cal N},h)$ and $({\cal N},h')$ be asymp\-to\-ti\-cal\-ly flat
Rie\-man\-nian
three-manifolds with
compact interior and finite mass, such that $h$ and $h'$ are $C^{1,1}$ and related via the
conformal rescaling $h' = \Omega^2 h$ with a $C^{1,1}$- function $\Omega >0$. Assume further that
there exists a non-negative constant $\beta$ such that the corresponding Ricci scalars satisfy
$R+ \beta \Omega^2 R' \geq 0$ everywhere. Then the corresponding masses satisfy $m+ \beta m'\geq 0$.
Moreover, equality holds iff both $({\cal N},h)$ and $({\cal N},h')$ are flat Euclidean spaces.
\end{lemma} 
{\it Proof.}
For the Ricci scalar $\widetilde R$ with respect to the metric
$\widetilde h = \Omega^{2\beta/(1 + \beta)} h$ we obtain, by standard formulas for conformal rescalings

\begin{eqnarray*}
(1 + \beta) \widetilde R =  \Omega^{- \frac{2 \beta}{1 + \beta}} (R + \beta \Omega^2 R') +
2 \beta(1 + \beta)^{-1} \Omega^{-2} \widetilde \nabla_{i} \Omega \widetilde \nabla^{i} \Omega.
\end{eqnarray*}
By the requirements of the lemma, $\widetilde h$ is AF and
$\widetilde R$ is non-negative. Hence, by virtue of the positive mass theorem
and by the relation $\widetilde m = (1 + \beta)^{-1} (m + \beta m')$ for the
masses we obtain the claimed results. $\hfill \Box$

\medskip

We can now easily prove our main result.
\begin{theorem}
Let $\Mg$ be a static, simply connected spacetime with a non-degenerate black hole
solving the Einstein-Maxwell-dilaton field equations in one of the following three cases:
\begin{itemize}
\item[(1)] The magnetic field vanishes.
\item[(2)] The electric field vanishes.
\item[(3)] The dilatonic coupling constant $\alpha$ is equal to
one.
\end{itemize}
Assume further that the mass and the charges satisfy 
${}^{\Phi}A_{-} \ne 0$ and ${}^{\Psi}A_{-} \ne 0$ (c.f.
(\ref{psiA}),(\ref{alA})). Then $\Mg$ must be a member of 
the spherically symmetric ``Gibbons-Maeda-'' family of solutions \cite{GM}.
\end{theorem} 
{\bf Remark.}
The condition that $\Mg$ is simply connected is used only to guarantee the global existence
of the electric and magnetic potentials $\phi$ and $\psi$. This condition fits
rather naturally to BH spacetimes. In concrete terms, if the exterior of the
BH is assumed to be globally hyperbolic, the ``topological censorship theorems'' of Chru\'sciel and
Wald \cite{CW} and Galloway \cite{GaGa} imply simply connectedness. Thus the conclusions
of the theorem hold for BH with a globally hyperbolic domain of outer communications.
\\ \\
{\it Proof.} We introduce the Ricci scalars ${}^{\Phi}{\cal R}$
and ${}^{\Psi}{\cal R}$ with respect to
${}^{\Phi} g_{ij}$ and ${}^{\Psi} g_{ij}$
(which should not be mixed up with ${}^{\Phi} R$ and ${}^{\Psi} R$), and
\begin{eqnarray*}
{}^{\Phi}E_{i} = {}^{\Phi}\Omega^{-1} \epsilon_{ab}\Phi^a \nabla_{i}\Phi^b,
\qquad
{}^{\Psi}E_{i} = {}^{\Psi}\Omega^{-1} \epsilon_{ab}\Psi^a \nabla_{i}\Psi^b,
\end{eqnarray*}
where $\epsilon_{12} = 1 = -\epsilon_{21}.$
We find that
\begin{eqnarray*}
\lefteqn{\psi  =  0: {}} \nonumber\\
& & {}
{}^{\Phi}\Omega^2~{}^{\Phi}{\cal R}+\alpha^2~{}^{\Psi}\Omega^2~{}^{\Psi}{\cal R}=
2~{}^{\Phi}\Omega^2~{}^{\Phi} g^{ij}~{}^{\Phi}E_{i}~{}^{\Phi}E_{j} +
2~{}^{\Psi}\Omega^2~{}^{\Psi} g^{ij}~{}^{\Psi}E_{i}~{}^{\Psi}E_{j},\\
\lefteqn{\alpha  =  1: {}} \nonumber\\
& & {}
{}^{\Phi}\Omega^2~{}^{\Phi}{\cal R}+ {}^{\Psi}\Omega^2~{}^{\Psi}{\cal R}=
2~{}^{\Phi}\Omega^2~{}^{\Phi} g^{ij}~{}^{\Phi}E_{i}~{}^{\Phi}E_{j} +
2~{}^{\Psi}\Omega^2~{}^{\Psi} g^{ij}~{}^{\Psi}E_{i}~{}^{\Psi}E_{j}.
\end{eqnarray*}
where the r.h. sides are manifestly non-negative.
Defining now $\beta = \alpha^2$ and the metrics $h = {}^{\Phi} g$, $h' = {}^{\Psi} g$ and
$\widetilde h = \Theta^2 g$ by
\begin{eqnarray*}
\lefteqn{\psi  =  0: {} \hspace{3cm}} \nonumber\\
& & {} \Theta^{1 + \beta} = {}^{\Phi} \Omega ~{}^{\Psi}\Omega^{\beta},\\
\lefteqn{\alpha  =  1: {}}, \nonumber\\
& & {} \Theta^2 = {}^{\Phi} \Omega ~{}^{\Psi} \Omega,
\end{eqnarray*}
we can apply the rigidity case of Lemma \ref{rescaling}, which yields that
$(\N,{}^{\Phi} g)$, $(\N,{}^{\Psi} g)$ and  $(\N,{} \widetilde g)$ are flat.
This also implies that ${}^{\Phi}\Omega_{\pm} = {}^{\Psi}\Omega_{\pm}$ and
hence $\Phi_{1} = \Psi_{1}$.
Furthermore, we have ${}^{\Phi}E_{i} =  {}^{\Psi}E_{i} = 0$,
which yields that
$a_{-1} \Phi_{-1} = a_{0} \Phi_{0}$ and $b_{-1} \Psi_{-1} = b_{0} \Psi_{0}$ for some
constants $a_{-1}$, $a_{0}$, $b_{-1}$ and $b_{0}$. Hence all potentials
$\Phi_A$ and $\Psi_A$ are functions of just a single variable. The following one is
particularly useful
\begin{eqnarray*}
 \Re^2 = \frac{A^2 (\Phi_1 + 1)}{4 (\Phi_1 - 1)}=
\frac{A^2 (\Psi_1 + 1)}{4 (\Psi_1 - 1)},
\end{eqnarray*}
where $A^2 = {}^{\Phi}A^{2} = {}^{\Psi}A^{2}$ (and
${}^{\Phi}A^{2}$ and ${}^{\Psi}A^{2}$ were defined before Lemma 3).
In fact, the field equations
(\ref{Riccpsi}),(\ref{Riccal}) and the flatness of $(\N,\widetilde g)$ imply that
$\widetilde \nabla_i  \widetilde \nabla_j \Re^2 = 2 \delta_{ij}$, and so $\Re$ coincides with
the standard radial coordinate in $\Bbb{R}^3$ (for details, c.f. the proof of Theorem 1 in \cite{MS}).
To obtain the form (\ref{GGPhiPsi}) we introduce the harmonic coordinate $\rho = \Re + A^2/4\Re$. 
$\hfill \Box$

\section{Harmonic maps} 
 
In this section we consider massless coupled harmonic maps in general, as introduced 
in Definition 3. Our aim is to obtain information on the possible conformal 
factors which are suitable for proving uniqueness of spherically symmetric BH following the method
of Bunting 
and Masood-ul-Alam. For this purpose we first study massless 
coupled harmonic maps where both $(\Sigma,g)$ and $\Upsilon$ 
are spherically symmetric. Any such map must be of the form $\Upsilon = \zeta \circ \lambda $, where 
$\zeta : I \subset \Bbb{R} \rightarrow \V$ is an affinely parametrized geodesic 
of $(\V,\gamma)$  and $\lambda : \Sigma \rightarrow \Bbb{R}$ is a spherically symmetric
 harmonic function on 
$\Sigma$ (see, e.g.\cite{BGM}).  Thus, spherically symmetric solutions are described by geodesics in the target space. 
However, in general not all geodesics of the target space correspond to a non-degenerate spherically
symmetric BH.
Let us put forward the following definition. 
 
\begin{definition} 
Let  $\V$ be the target space of a coupled harmonic map. We define $\V_{BH} \subset \V$ as 
$\V_{BH} = \{x \in \V |$there exists a spherically symmetric,
 non-degenerate black hole spacetime whose defining geodesic in the target space 
passes through $x \}$. 
\end{definition} 
{\bf Remark.} 
In this section we will assume that the condition of AF (implicit in our definition 
of a BH in Sect. 2) restricts the values of the matter fields and the norm of the static Killing 
vector to specific values at infinity (perhaps after suitable shifts and/or
rescalings of the fields). For example, in the case of EMD we have chosen $V=\tau=1$, $\phi=\psi=0$
 at infinity. 
The corresponding point in the target space will be denoted by $p_{\infty}$. Thus, every 
point $x \in \V_{BH}$ can be joined to $p_{\infty}$ by at least one geodesic giving rise to
a non-degenerate, spherically symmetric BH. We will denote any such geodesic
by $\zeta_x(s)$ and we will fix the (affine) parametrization uniquely by demanding (without loss
of generality) $\zeta_x(0)=p_{\infty}$, $\zeta_x(1)=x$. Notice that this condition restricts the harmonic 
function $\lambda$ appearing in $\Upsilon = \zeta_x \circ \lambda$ to satisfy $\lambda =0$ at  infinity 
in $\Sigma^{\infty}$. A geodesic passing through $p_{\infty}$ will qualify as a defining geodesic
for a spherically symmetric BH provided several conditions are met on the BH boundary. In particular, it is necessary that the 
geodesic reaches $V=0$ (i.e. the horizon of the BH) at an infinite value of the affine parameter and that the rest
of fields remain finite there. A more detailed description of the necessary and
sufficient conditions in the case of target spaces which are symmetric spaces can be found in \cite{BGM}. 

Notice that the subset $\V_{BH}$ need not be a submanifold (although this happens to be the case in all 
explicit cases known to us). In particular, it could happen that two or more geodesics 
connecting $x$ and $p_{\infty}$ define spherically symmetric BH. However, such geodesics  would define different 
BH solutions so that we can still associate to every solution one geodesic in a unique way. 

\medskip 

In Theorem 2 below we shall determine explicitly and uniquely the conformal factors on $\V_{BH}$ for 
which the method of Bunting and Masood-ul-Alam has a chance to work. In the following, we shall be dealing
with objects on $\Sigma$ which are the pull-backs of 
objects on $\V$ under $\Upsilon$. In order to avoid cumbersome notation we shall 
use the same symbol for both objects. The precise meaning should become clear 
from the context. 
 
\begin{theorem} 
Consider a massless coupled harmonic map with target space $(\V,\gamma)$ and let 
$\Omega_{\pm}$ be positive, $C^2$ functions $\Omega_{\pm} : \V 
\rightarrow \Bbb{R}$ with the following properties 
\begin{itemize} 
\item[(1)] For any sphe\-ri\-ca\-lly sym\-me\-tric, non-degenerate,
sta\-tic black ho\-le 
$(\Sigma_{sph}, g_{sph})$ the metric $\Omega_{\pm}^2 g_{sph}$ is locally
flat. 
\item[(2)] $(\Sigma^{\infty}_{sph},(\Omega_{+})^2 g_{sph})$ is 
asymptotically flat and $(\Sigma^{\infty}_{sph},(\Omega_{-})^2 g_{sph})$ 
admits a one-point compactification of infinity. 
\end{itemize} 
Then, $\Omega_{\pm}$ must take the following form on $\V_{BH}$ 
\begin{eqnarray} 
\Omega_{+} (x) = \cosh^2 
\sqrt{\frac{\dot{\zeta}^a(x) \dot{\zeta}_a(x)}{8}}
, \hspace{4mm} 
\Omega_{-}(x) = \sinh^2 
\sqrt{\frac{\dot{\zeta}^a(x) \dot{\zeta}_a(x)}{8}} 
\hspace{4mm} \forall x \in \V_{BH} 
\label{explicit}
\end{eqnarray} 
where $\dot{\zeta}^a(x)$ is the tangent vector at $x$ 
of the geodesic $\zeta_x(s)$ in $(\V,\gamma)$ 
defining the spherically symmetric black hole $(\Sigma_{sph},g_{sph})$.

Conversely, for any spherically symmetric static black hole 
$(\Sigma_{sph},g_{sph})$, the metrics $\Omega^2_{\pm} g_{sph}$, 
with $\Omega_{\pm}$ given by (\ref{explicit}) are locally flat.
 
\label{Omegapm} 
\end{theorem} 
{\it Proof.}  From the coupled harmonic map equations (\ref{chm1}), (\ref{chm2}) and from the
behaviour of the Ricci scalar under a conformal rescaling $g' = \Omega^2 g$, where 
$\Omega$ is a function $\V \rightarrow \Bbb{R}$, we find  
\begin{eqnarray} 
\Omega^2 R' = \left(\gamma _{ab}- 
4 \frac{D_a D_b\Omega}{\Omega} 
+2 \frac{D_a\Omega D_b \Omega}{\Omega^2}\right) 
\nabla_i\Upsilon^a \nabla^i\Upsilon^b 
\label{conftrans} 
\end{eqnarray} 
where $D$ is the covariant derivative on $\V$. 
Let $x \in \V_{BH}$ and $\zeta_x$ be the geodesic in $(\V,\gamma)$ 
giving rise to the spherically symmetric BH $(\Sigma_{sph},g_{sph})$. 
Applying (\ref{conftrans}) to this solution and using 
Condition (1) we obtain, with $\Omega_{\pm} = 
(\sigma^{\pm})^2$, 
\begin{eqnarray} 
0 = \left (\gamma_{ab} - 8 
\frac{D_{a} D_{b} \sigma^{\pm}}{\sigma^{\pm}} 
\right ) \dot{\zeta}^a (\lambda) 
\dot{\zeta}^b (\lambda) \nabla_i \lambda \nabla^i \lambda. 
\label{EqSpher} 
\end{eqnarray} 
Defining ${\tilde{\sigma}}^{\pm} = \sigma^{\pm} \circ \zeta_x$, equation 
(\ref{EqSpher}) becomes, after using the fact that $\zeta_x$ is a geodesic, 
\begin{eqnarray} 
\frac{ d^2 {\tilde{ \sigma}}^{\pm}(s)}{ds^2} = 
\frac{N(x)}{8} {\tilde{\sigma}}^{\pm}(s) 
\label{ODE} 
\end{eqnarray} 
where $N(x) = \dot{\zeta}^a (x) \dot{\zeta}_a (x)$. Condition (2) imposes, 
first of all, that $\Omega_{+}(p) = 1$ and $\Omega_{-}(p) =0$, or, equivalently, 
\begin{eqnarray} 
{\tilde{\sigma}}^{\pm}(0) = 1/2 \pm 1/2. \label{boundcondODE} 
\end{eqnarray} 
Furthermore, under a conformal rescaling $g' = \sigma^4 g$, 
the mass changes according to 
\begin{eqnarray} 
\label{mass} 
m - m' = \frac{1}{2 \pi} \int_{S^{\infty}} \nabla_i \sigma dS^i, 
\end{eqnarray} 
where $S^{\infty}$ stands for the sphere at infinity in 
$\Sigma^{\infty}$. For the conformal factor $\sigma^{+}$, the 
right-hand side is zero because the metric $g_{sph}$ has vanishing 
mass and $(\sigma^{+})^4 g_{sph}$ is flat. 
Similarly, for $\sigma^{-}$, infinity is compactified to a point 
and so the right-hand side must also vanish (the sphere at 
infinity becomes a point). Let $S_r$ be a sphere or radius $r$ in 
$\Sigma_{sph}^{\infty}$ ($r$ sufficiently large). Then 
(\ref{mass}) implies 
\begin{eqnarray} 
0 =  \lim_{r \rightarrow \infty} \int_{S_r} 
 \left . \frac{d {\tilde{\sigma}}^{\pm}}{ds} 
\right |_{s = \lambda} \nabla_i \lambda 
dS^i.
\label{spherinf} 
\end{eqnarray} 
A trivial analysis of the Laplace equation for spherically symmetric functions in a 
spherically symmetric, AF spacetime shows that 
$\nabla_i \lambda = \mbox{O} (r^{-2})$. Thus, (\ref{spherinf}) implies 
$\frac{d {\tilde{\sigma}}^{\pm}}{ds} |_{s = 0} =0$. 
The unique solution of the ODE (\ref{ODE}) fulfilling this initial 
condition and (\ref{boundcondODE}) is 
\begin{eqnarray*} 
\sigma^{+}(x) =  \cosh \sqrt{\frac{N(x)}{8}}  \hspace{5mm} 
\sigma^{-}(x) = \sinh \sqrt{\frac{N(x)}{8}},
\hspace{5mm} \forall x \in \V_{BH} 
\end{eqnarray*} 
and the first part of the Theorem follows. 

In order to prove the converse, we need to show that
the Ricci tensor $R^{\pm}_{ij}$ of the conformally rescaled metric 
$\Omega_{\pm}^2 g_{sph}$ vanishes for any spherically symmetric BH.
From the previous results we know that the Ricci scalar $R^{\pm}$
vanishes. From spherical symmetry, it follows that we only need to check whether the
radial component of $R^{\pm}_{ij}$ vanishes, i.e. $R^{\pm}_{ij} \nabla^{i} \lambda \nabla^{j} \lambda =0$. 
From the coupled harmonic map
equations (\ref{chm1}),(\ref{chm2}) and from spherical symmetry it follows,
at any point $y \in \Sigma_{sph}$ and for any function $F : \V \rightarrow \Bbb{R}$, 
\begin{eqnarray}
\nabla_i \nabla_j F (y) = - \frac{1}{4} \left .
\frac{d G}{d \lambda} \right |_y
\dot{\xi}^a(x)
\partial_a F(x) \left . \left (h_{ij} - 2 n_{i} n_{j} \right ) \right |_{y} + \nonumber
\hspace{2cm} \\
+ \dot{\xi}^a(x) \dot{\xi}^b(x) 
D_a D_b F (x) \left . \left ( \nabla_i \lambda \nabla_j \lambda \right ) \right )|_{y} 
\label{derderF}
\end{eqnarray}
where $x = \Upsilon (y)$, $G = \nabla_i \lambda \nabla^i \lambda$, $n_i$ is the unit
radial normal of $(\Sigma_{sph}, g_{sph})$ (i.e. 
$n_i = G^{-1/2} \nabla_i \lambda$ wherever $G \neq 0$) and $h_{ij}$ is the projector orthogonal
to $n_i$. Using (\ref{derderF}) for $F = \sigma^{\pm}$, (\ref{chm2}), (\ref{EqSpher}) 
and the transformation law for the Ricci tensor under conformal rescalings we obtain

\begin{eqnarray}
\left . R^{\pm}_{ij} \nabla^i \lambda \nabla^j \lambda \right |_{y} = 4 G^2(y) \left (
- \left . \frac{1}{4 G}  \frac{d G}{d \lambda} \right |_{y} 
\left . \frac{D_a \sigma^{\pm}}{\sigma^{\pm}} \right |_{x}
\dot{\xi}^a(x) +  \left . \frac{D_a D_b \sigma^{\pm}}{\sigma^{\pm}} \right |_{x}
\dot{\xi}^a (x) \dot{\xi}^b (x) + \right . \nonumber \\
 \left . 
 + \left . \frac{D_a \sigma^{\pm} D_b \sigma^{\pm}}{{\sigma^{\pm}}^2} \right |_{x}
 \dot{\xi}^a (x) \dot{\xi}^b (x)
\right ). \label{radial}
\end{eqnarray}
Next we need to evaluate $G$. Again, from spherical symmetry and the
coupled harmonic map equations it follows that $R_{\theta\theta} =0$.
This determines the metric $g_{sph}$ modulo two constants. Integrating the
spherically symmetric Laplace equation on this background, the following expression is obtained
\begin{eqnarray*}
-\frac{1}{4 G} \frac{d G}{d \lambda} (y) = \sqrt{\frac{N(x)}{2}} \mbox{cotanh}
\sqrt{\frac{N(x)}{2}}.
\end{eqnarray*}
Inserting this into (\ref{radial}) and using the explicit form for $\sigma^{\pm}$ we obtain 
$R^{\pm}_{ij} \nabla^i \lambda \nabla^j \lambda =0$, which proves the
claim. $\hfill \Box$.\\ \\
{\bf Remark.} 
Notice that the proof of the theorem holds for {\it any} geodesic passing through $x \in \V_{BH}$
defining a spherically symmetric BH. Thus, if there exists a point $x \in \V_{BH}$
with two or more geodesics defining a BH, then the proof of Bunting and Masood-ul-Alam can 
work {\it only} provided the square distance function $N(x) = \dot{\xi}^a(x) \dot{\xi}_a(x)$ takes the
same value on each of these geodesics (so that the conformal factors $\Omega_{\pm}$ are well-defined). 
This should be viewed as a restriction on the set of coupled harmonic map models for which 
the method of \cite{BM} may work. As mentioned above, all known models
have the property that the ``BH''-geodesic passing through $x \in \V_{BH}$ is unique.
 
\medskip 
 
We conclude with a discussion of the results of this section, with the EMD case serving as example. 
 
The crucial step in the uniqueness proof of Bunting and Masood-ul-Alam \cite{BM} 
is to define appropriate conformal factors on the target space which rescale the 
spherically symmetric BH to flat space and which have the appropriate asymptotic
behaviour. In Theorem 2 above, 
we have used these properties as guiding principles to define, for general
coupled harmonic maps, unique functions $\Omega_{\pm}:
\V_{BH} \rightarrow \Bbb{R}$ on a certain subset $\V_{BH}$ of the 
target space $\V$. These ``candidates'' for conformal factors would 
be perfectly suited for a uniqueness proof if they could 
(i) be extended suitably
to the whole target space if $\V_{BH}$ is smaller than the whole target space
and (ii) be shown to be positive, having the right behaviour at infinity and at the horizon for 
every coupled harmonic map (without the assumption of spherical symmetry) 
and with rescaled Ricci scalar being non-negative. 
 
In vacuum and in EM theory, $\V_{BH}$ coincides 
with the target space and the theorem above yields unique conformal factors 
(which coincide with the ones used in \cite{WS2,AM1}).
The uniqueness obtained here is remarkable for the following 
reason. It is clear that there are infinitely many possibilities 
to combine the potentials $V$ and $\phi$ to factors which rescale 
the Reissner-Nordstr\"om metric to the flat one, and which have the 
right boundary conditions. In fact, we can just take $\Omega$ to be 
either a suitable function only of $V$ or only of $\phi$. However, such 
conformal factors would in general depend explicitly on the mass 
$M$ and the charge $Q$ of the solution and therefore would not be
functions on $\V$ as required in Theorem 2. In this situation there 
is little hope for proving  that the rescaled metrics yield 
non-negative rescaled Ricci scalars. For this reason we believe that 
the assumption that the conformal factors depend only on the target 
space variables is quite reasonable in general. 
 
If $\V_{BH}$ is smaller than $\V$, the factors obtained from 
Theorem 2 on $\V_{BH}$ have to be extended to $\V$, which involves 
``guesswork''. In each of the three special cases of 
EMD theory considered above, $\V_{BH}$ has codimension one in $\V$, and guessing 
the ``right'' 
factors is easy (with or even without using Theorem 2) after the simple 
structure of the symmetry group of $\V$ is recognized. 

 In the general EMD case with $\alpha \ne 1$, the isometry group of $\V$ is 
aff(1) $\times$ aff(1) where aff(1) is the group of affine motions of the line. 
However, we are not aware of explicit forms of the spherically symmetric BH and 
we have no knowledge of $\V_{BH}$ (not even of its dimensionality).
Thus we have here neither a systematic way for determining, nor even for guessing good 
conformal factors.  In fact, it is plausible that solving the general EMD case could give 
interesting clues on how to extend the conformal factors off $\V_{BH}$ for general harmonic maps.

\bigskip

{\Large\bf Acknowledgement.} We are grateful to the referee for
pointing out errors in the draft and for suggesting improvements, and to
Helmuth Urbantke for helpful discussions.

\end{document}